\documentclass[cameraready]{vgtc}             
\onlineid{1027}

\graphicspath{{./figs/}{./figsdraft-new/}} 

\usepackage{amsfonts}
\usepackage{amsmath}
\usepackage{multirow}

\vgtccategory{Research}
\vgtcinsertpkg

\usepackage{tabu}      
\usepackage{booktabs}

\usepackage{times}                     

\usepackage{mathptmx}                  

\newcommand{\Mspace}{\mathbb{M}}
\newcommand{\Rspace}{\mathbb{R}}

\newcommand{\Ind}{\mathrm{I}}

\newcommand{\floor}[1]{\left\lfloor #1 \right\rfloor}
\newcommand{\ceil}[1]{\left\lceil #1 \right\rceil}

\newcommand{\para}[1]        {\vspace{1pt}\noindent{\textbf{#1}}}

\usepackage{graphicx} 
\usepackage{amsthm}
\usepackage[noend]{algorithmic}
\usepackage{algorithm}
\newtheorem{theorem}{Theorem}

\usepackage{subcaption}
\usepackage{enumitem}

\usepackage{makecell}
\newcommand{\CellWithForceBreak}[2][c]
{\begin{tabular}[#1]{@{}c@{}}#2\end{tabular}}

\newcommand{\update}[1]{#1}
\newcommand{\modify}[1]{#1}

\DeclareMathAlphabet{\mathcal}{OMS}{cmsy}{m}{n}
\SetMathAlphabet{\mathcal}{bold}{OMS}{cmsy}{b}{n}
\usepackage{mathtools, nccmath}

\usepackage[left=0.75in, right=0.75in, top=0.8in, bottom=0.8in]{geometry} 
\usepackage{caption} 

\title{Critical Point Extraction from Multivariate Functional Approximation}

\author{Guanqun Ma \thanks{e-mail: guanqun.ma@utah.edu} \\
\scriptsize University of Utah
\and David Lenz \thanks{e-mail: dlenz@anl.gov} \\
\scriptsize Argonne National Laboratory
\and Tom Peterka \thanks{e-mail: tpeterka@mcs.anl.gov} \\
\scriptsize Argonne National Laboratory
\and Hanqi Guo \thanks{e-mail: guo.2154@osu.edu} \\
\scriptsize Ohio State University
\and Bei Wang \thanks{\vspace{-60pt} e-mail: beiwang@sci.utah.edu} \\
\scriptsize University of Utah}

\abstract{
Advances in high-performance computing require new ways to represent large-scale scientific data to support data storage, data transfers, and data analysis within scientific workflows. 
Multivariate functional approximation (MFA) has recently emerged as a new continuous meshless representation that approximates raw discrete data with a set of piecewise smooth \update{functions}. 
An MFA model of data thus offers a compact representation and supports high-order evaluation of values and derivatives anywhere in the domain. 
In this paper, we present CPE-MFA, the first critical point extraction framework designed for MFA models of large-scale, high-dimensional data. 
CPE-MFA extracts critical points directly from an MFA model without the need for discretization or resampling. 
This is the first step toward enabling continuous implicit models such as MFA to support topological data analysis at scale.

}

\begin{document}

\maketitle

\section{Introduction}
\label{sec:introduction}

Advances in high-performance computing (HPC) require new ways to represent large-scale scientific data to support data storage, data transfers, and data analysis.  
To that end, there has been a growing interest in replacing discrete data with continuous, high-order, and differentiable representations, such as functional models and implicit neural networks~\cite{SitzmannMartelBergman2020,SitzmannZollhoferWetzstein2019,MartelLindellLin2021,MildenhallSrinivasanTancik2020}, to enhance scientific workflows. 

Among the functional models, multivariate functional approximation(MFA)~\cite{peterka2018foundations,peterka2022multivariate} has recently emerged as a new continuous meshless representation of discrete data. 
It approximates the raw discrete data by a set of piecewise smooth polynomial functions, and supports high-order evaluation of values and derivatives anywhere in the domain. 
An MFA model offers a compact representation as it consumes less storage space than the original discrete data, and serves as a surrogate in supporting spatiotemporal analysis in the continuous domain.

MFA is a technology supported by the United States Department of Energy (DOE) under the SciDAC RAPIDS Institute~\cite{rapids2} and has been applied to large-scale projects in high-energy physics~\cite{HepOnHPC} and climate science~\cite{seahorce}. MFA was originally proposed as a tool for modeling structured scientific datasets~\cite{peterka2018foundations}, and has been expanded to handle complex unstructured point clouds~\cite{lenz2023customizable, lenz2022adaptive} and operates in distributed parallel environments \cite{mahadevan2023domain,sun2024mfa}. It can also be used as an intermediate representation for high-quality volume rendering~\cite{sun2023scalable}.

However, an unexplored area of research is to utilize continuous implicit models such as MFA for topological data analysis (TDA) and visualization. 
A fundamental step to recovering topology from an MFA model is to extract its critical points. 
In Morse theory, critical points correspond to topological changes in scalar fields and they play a crucial role in TDA~\cite{edelsbrunner2022computational}. 
In this paper, we present CPE-MFA, the \emph{first}  critical point extraction framework designed for MFA models of large-scale, high-dimensional data. Our method represents the first approach to enable topological feature extraction from implicit continuous models (such as MFA). A standard MFA model is designed as a forward model to evaluate variables at query locations. In contrast, our CPE-MFA framework facilitates inverse evaluations for such a forward model, broadening the applicability of MFA in TDA.

In MFA, piecewise functional approximations replace discrete data with linear combinations of \emph{basis functions} (e.g.,~nonuniform rational B-spline functions or NURBS) and a set of reference points called \emph{control points}~\cite{peterka2018foundations}. 
The places where the piecewise functions meet are known as \emph{knots}, and the interval between knots are \emph{knot spans} (or just \emph{spans}). 
The function value in a span depends on a set of control points. 
Based on the known control points, we can first bypass spans devoid of critical points. 
We then employ Newton's method to extract critical points in the remaining spans. 
Since each piecewise functional is a polynomial, we could use its closed-form derivative to obtain the critical points, greatly reducing the complexity of Newton's method. 
Finally, we introduce spatial hashing to remove duplicated copies of critical points. Our contributions include: 
\begin{itemize}[noitemsep]
\item We introduce a novel framework, called CPE-MFA, for critical point extraction from MFA models. Our framework extracts critical points directly from an MFA model without the need for discretization or resampling. 
\item We propose multiple strategies to enhance the efficiency during critical point extraction, including multithreading, bypassing specific spans devoid of critical points, and removing duplicates using spatial hashing. 
\item We demonstrate the efficacy of our framework across multiple scientific datasets. 
\end{itemize}
\vspace{-8pt}
Our work is the first step toward enabling continuous implicit models such as MFA to support topological data analysis at scale. 

\section{Related Work}
\label{sec:related-work}

We review relevant work on MFA and critical point extraction.
 
\para{Multivariate Functional Approximation,}~or MFA, is a method for representing a scientific dataset with a continuous B-spline function for the purposes of analysis and visualization~\cite{peterka2018foundations, peterka2022multivariate, lenz2023customizable}. MFA may be considered a form of scattered data approximation (SDA), a field of mathematics concerned with constructing continuous functions that approximate a spatial dataset~\cite{wendland2004scattered}. Numerous SDA methods have been developed over the years, with popular functional approximations based on wavelets~\cite{jansen2005second}, radial basis functions~\cite{majdisova2017radial}, and splines~\cite{deBoor2001guide}. 

MFAs are built on geometric basis functions, specifically, B-splines. B-splines and their generalization, nonuniform rational B-splines or NURBS are smooth, flexible curves widely used in modeling and visualization software~\cite{rogers2001introduction, lin2018geometric}. Before MFA, Martin and Cohen proposed a model using NURBS to represent data in 2D and 3D~\cite{martin2001representation}. This work was extended by Martin et al. \cite{martin2008volumetric} to parameterize 2D triangular and 3D tetrahedral data with tensor product splines.  MFA is an extension of this method to model data with any number of dimensions. 

Some visualization algorithms have been built upon geometric functional representations in the past.  In 1997, Park and Lee~\cite{park1997high} utilized a high-dimensional trivariate NURBS representation to visualize fluid flow data. In 2001, Martin and Cohen~\cite{martin2001representation} constructed isosurfaces and achieved ray tracing using NURBS. More recently, Sun et al.~\cite{sun2023scalable, sun2024mfa}  developed techniques for scalable, interactive volume visualization of MFA representations. Their techniques  produced high quality renders with moderate time complexity, and introduced fewer visual artifacts than traditional local filtering techniques.

\para{Critical point extraction.}
\update{TDA} utilizes topological descriptors and provides robust feature extraction techniques for large-scale scientific data; see \cite{HeineLeitteHlawitschka2016,YanMasoodSridharamurthy2021} for surveys.   
These techniques have been applied to diverse research fields such as chemistry~\cite{olejniczak2020topological,bhatia2018topoms}, astrophysics~\cite{sousbie2011persistent,shivashankar2015felix}, and biomedical imaging~\cite{anderson2018topological,bock2018topoangler}. 
\update{Critical points may correspond to atoms in chemistry \cite{bhatia2018topoms}, galaxy clusters in astrophysics \cite{sousbie2011persistent,shivashankar2015felix}, and serve as good seed points for neuron reconstruction in biomedical imaging \cite{chen2021spherical}.}
For scalar field data, critical points are part of the foundational structures of topological descriptors, such as merge trees, contour trees~\cite{carr2003computing,smirnov2020triplet}, Reeb graphs~\cite{reeb1946points}, and Morse-Smale complexes~\cite{edelsbrunner2001hierarchical,edelsbrunner2003morse}. 
Morse theory~\cite{milnor1963morse} plays an important role in the study of critical points. 
Forman~\cite{forman2002user} generalized the theory from the smooth setting to the discrete setting, referred to as the discrete Morse theory, making it practical for studying discrete data. 
The challenge of extracting critical points for discrete data arises due to the absence of proper differentiability. 
Banchoff~\cite{banchoff1970critical} characterized piecewise-linear (PL) critical points for an input PL scalar field defined on PL manifolds, based on lower- and upper-links of vertices. 
In particular, a vertex is \emph{regular} if both its lower- and upper-links are simply connected, and \emph{critical} otherwise.  
While there are various notions of PL critical points in the literature, they are shown to be equivalent, and correspond to discrete Morse cell in discrete Morse theory~\cite{FugacciLandiVarl2020}. 

Recently, Vidal et al.~\cite{vidal2021progressive} introduced a progressive method to extract critical points from PL functions defined on triangulated meshes. 
Their method utilized the fast identification of topologically invariant vertices using a hierarchical data representation. 
We use this method in~\cref{sec:results} as a reference to evaluate our framework. 

There are numerous studies on the numerical extraction of critical points, especially from a vector field perspective. 
Helman and Hesselink \cite{helman1989automated} first located isolated non-degenerate critical points of vector fields based on the Jacobian.
A number of methods focused on solving a system of linear equations in a PL vector field, for each cell in a triangulated domain \cite{li2006representing, lavin1998feature}. 
Skala and Smolik \cite{skala2019new} approximated discrete vector fields with radial basis functions (RBFs). In particular, based on a vector field represented as an RBF approximation, they constructed a function related to the speed of a particle and extracted its critical points.

\section{Technical Background}
\label{sec:background}

\subsection{Multivariate Functional Approximation}
\label{sec:MFA}

At their core, MFA models are tensor-product B-spline functions that serve as an approximation to a dataset. In this section, we provide an overview of the basic definitions and constructions necessary to describe B-spline models for scientific data. A thorough presentation on the fundamental theory of B-splines can be found in the books by de Boor~\cite{deBoor2001guide} and Piegl and Tiller~\cite{piegl1997nurbs}.

\para{B-spline curves.}
First, we consider the 1-dimensional case to illustrate the main components of a B-spline. Consider a set of point locations $\{u_0, \ldots, u_{m-1}\}\subset [0,1]$ with a value $f_i$ associated to each point.\footnote{We assume the input are re-scaled to lie in $[0,1]$.}
A best-fit B-spline curve is a function $F:\mathbb{R}\to \mathbb{R}$ such that $F(u_i) \approx f_i$ for all indices $i$, subject to certain conditions that we will define shortly.

\begin{figure}[!ht]
\vspace{-2mm}
\centering
\includegraphics[width=0.9\columnwidth]{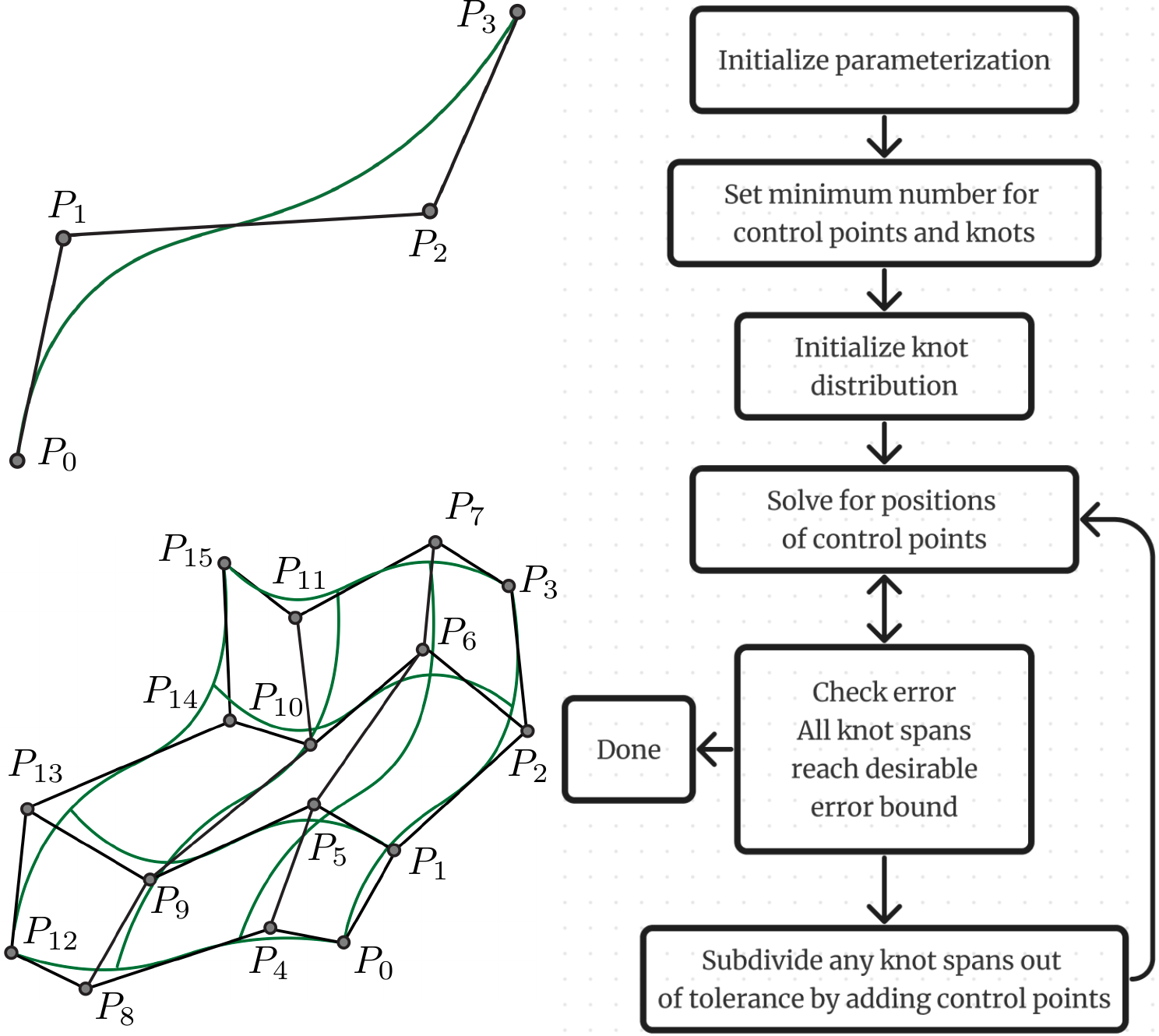}
\vspace{-3mm}
\caption{Left: a 1-dimensional B-spline curve (top), and a 2-dimensional B-spline surface (bottom). $P_i$ are control points, control meshes are in black, approximated curve/surface are in green. Right: an overall MFA model pipeline reproduced from~\cite[Fig.~1.4]{peterka2022multivariate}.}
  \label{fig:pipeline}
  \vspace{-4mm}
\end{figure}

A B-spline curve of \emph{degree p} is a piecewise-polynomial function with $p-1$ continuous derivatives, where the pieces are polynomials of degree $p$. The points where the curve transitions between polynomial pieces are called \emph{knots}, and the interval between two subsequent knots is a \emph{knot span}. The overall shape of the B-spline is determined by the location of geometric \emph{control points} scattered throughout the domain. The B-spline curve smoothly follows the polyline given by the control points, but does not coincide with the control points. Intuitively, it can be seen that B-splines with more control points can twist and bend with greater flexibility than those with fewer control points. 
\cref{fig:pipeline} (left) shows examples of a 1-dimensional B-spline curve (top) and a 2-dimensional B-spline surface. 

Throughout this paper, we will denote the degree of a B-spline by $p$. The set of control points is $\{P_j\}_{j=0}^{n-1}$, and the set of knots is denoted $\{t_j\}_{j=0}^{n+p}$. We note that the number of knots and control points in a B-spline are closely linked: a degree-$p$ spline with $n$ control points must have $n+p+1$ knots~\cite{deBoor2001guide}. Mathematically, B-splines may be described as linear combinations of B-spline basis functions, where the coefficient on each basis function is a control point. Namely, 
\begin{equation}
    F(u) = \sum_{j=0}^{n-1} N_{j,p}(u) P_j,
\end{equation}
where the functions $N_{j,p}$ are bump functions that are nonzero in the subinterval $[t_j, t_{j+p+1}]$ and zero elsewhere. For a full description of the basis functions we refer to the canonical text by de Boor~\cite{deBoor2001guide}.

The purpose of MFA is to construct the B-spline that minimizes the root mean squared (RMS) error between a set of values and the spline's approximation to those values. Given a set of point-value pairs $\{u_i, f_i\}$, a degree $p$, and a set of knot locations $\{t_j\}$, the best-fit B-spline is the B-spline function $F$ that solves the minimization problem:
\useshortskip
\begin{equation}
    \min_F \left(\frac{1}{m}\sum_{i=0}^{m-1} |f_i - F(u_i)|^2 \right)^{1/2}.
\end{equation}

MFA can also construct a B-spline model using an adaptive feedback loop.
\cref{fig:pipeline} (right) reproduces the pipeline to construct an MFA model from the raw input data with an adaptive refinement algorithm. 
MFA begins by rescaling input data points to the interval $[0,1]$ and initializing a knot distribution with a small number of control points. The loop begins by computing the best-fit B-spline given the coarse distribution of knots, and checking if the desired error tolerance is achieved. If not, MFA repeatedly adds new knots, computes a new best-fit B-spline over the new knots, and checks the error again. The process ends when the resultant B-spline has the desired pointwise errors, or a maximum number of knots are added.

\para{B-Splines in higher dimensions.}
The preceding discussion can be extended to 2- and higher-dimensional data with tensor-product B-splines. In a tensor-product B-spline, the basis functions are still bump functions and are formed by multiplying 1-dimensional B-spline basis functions from each dimension together. 
For a 2-dimensional case, a B-spline of degree $p$ has two sets of knots $\{t^{(1)}_{j_1}\}_{j_1=0}^{n_1+p}$ and $\{t^{(2)}_{j_2}\}_{j_2=0}^{n_2+p}$ for the $x$ and $y$ dimensions, respectively. 
The 2-dimensional basis functions are of the form $N_{j_1,j_2}(u,v) = N^{(1)}_{j_1,p}(u)N^{(2)}_{j_2,p}(v)$. Consequently, the mathematical expression for a 2-dimensional tensor-product B-spline is as follows, visualized in~\cref{fig:pipeline} (left bottom), 
\begin{equation}
\label{eq:B_spline_2d}
    F(u,v) = \sum_{j_1 = 0}^{n_1-1} \sum_{j_2 = 0}^{n_2-1}N^{(1)}_{j_1,p}(u)N^{(2)}_{j_2,p}(v) P_{j_1,j_2}.
\end{equation}
For the general $d$-dimensional case, the form of a tensor-product B-spline is
\begin{equation}
  \begin{aligned}
      F(u_1,&\cdots, u_d)=\\
      &\sum_{j_1}\cdots \sum_{j_d} N^{(1)}_{j_1, p}(u_1)\cdots N^{(d)}_{j_d, p}(u_d) P_{j_1,\cdots, j_d}.
  \end{aligned}
  \label{eq:MFA}  
\end{equation}
In one dimension, the term ``knot span'' refers to the subinterval $[t_j, t_{j+1}]$ between any two adjacent knots. In higher dimensions, a knot span refers to the tensor product of 1-dimensional knot spans. Thus a $d$-dimensional span would be a region like 
\[[t^{(1)}_{j_1}, t^{(1)}_{j_1+1}]\times\cdots\times[t^{(d)}_{j_d}, t^{(d)}_{j_d+1}].\]

\para{Strong convex hull property.}
Unlike other continuous models for scientific data, B-splines have a useful property that can be leveraged to dramatically accelerate the extraction of critical points.  The idea behind the \emph{strong convex hull property} of B-splines is that, within a given knot span, the values of the spline function are entirely contained within the convex hull of the neighboring control points. More specifically, we can say the following:

\begin{theorem}[Strong Convex Hull Property]
\label{theorem:convex-hull}
If $(u_1,\cdots, u_d)$ is a point in the span $[t^{(1)}_{j_1},t^{(1)}_{j_1+1}]\times\cdots\times[t^{(d)}_{j_d},t^{(d)}_{j_d+1}]$, then $F(u_1,\cdots, u_d)$ lies in the convex hull defined by control points $P_{h_1,\cdots, h_d}$, where $j_l \leq h_l \leq j_l + p$, for $1\leq l \leq d$. 
\end{theorem}

\subsection{Critical Point Extraction}

\para{Newton's Method,}~also known as the Newton-Raphson method, is a widely used numerical algorithm to find the roots of a differentiable function. 

Suppose $f: \Rspace \to \Rspace$ is a differentiable scalar function. 
Starting with an initial guess of a value $x_0$, the iterative process is repeated as 
\begin{equation}
  x_{n+1}=x_n-\nabla f(x_n)^{-1}f(x_n), 
\end{equation}
where $n$ is the iteration number, $x_n$ is the current estimate of the root, and $\nabla f(x_n)$ is the gradient of $f$ at $x_n$. The iteration stops after step $n$ when $||\nabla f(x_{n})||$ is smaller than a predefined threshold $\varepsilon$. 
 
MFA is a continuous polynomial function, and its high-order derivatives can be calculated analytically. 
Suppose $f: \Rspace^d \rightarrow \Rspace$ is a general differentiable function defined on a $d$-dimensional domain. 
Finding the critical points of MFA is equivalent to solving the equation $\nabla f(x) = 0$. Newton's method fits well in this scenario. 
Starting with an initial guess, a $d$-dimensional vector $x_0$,  the iterative formula is
\begin{equation}
  x_{n+1}=x_n-H(x_n)^{-1}\nabla f(x_n).
\end{equation} 
Here, $x_n$ is a $d$-dimensional vector at the $n$th iteration, $H(x_n)^{-1}$ is the inverse of the Hessian matrix of $f$ at $x_n$, $\nabla f(x_n)$ is the gradient of the function $f$ at $x_n$. 
Iteration terminates when $||\nabla f(x_n)|| \leq \varepsilon$.

\para{Critical points.}
Geometrically, the goal of studying critical points is to extract features surrounding them that help scientists understand and represent data for downstream analysis (e.g.,~\cite{allili2011detecting}). 
Critical point detection is the identification of extrema and saddles that assists in understanding the topology of a  scalar field. 

Morse theory \cite{milnor1963morse} is a mathematical tool to detect and classify critical points. 
Given a function $f: \Mspace \to \Rspace$ defined on a smooth manifold $\Mspace$, the gradient $\nabla f$ vanishes at a critical point $p$. 
A critical point $p$ is \emph{non-degenerate} if its Hessian matrix $H(p)$ is not singular.  
A Morse function requires all its critical points to be non-degenerate and have distinct function values. 
The index $\lambda(p)$ of a critical point $p$ is defined by the number of negative eigenvalues of its Hessian $H(p)$: local maxima have  zero negative eigenvalues, local minima have all negative eigenvalues, and saddles have some (but not all) negative eigenvalues.
Extraction critical points in a PL setting is described in the supplement.

\section{Method}
\label{sec:method}
Our CPE-MFA framework consists of three main stages. 
First, we bypass spans devoid of critical points and work with the remaining valid ones. 
Second, we utilize Newton's method to extract critical points from each valid span. 
Since each piecewise functional is a polynomial, we use its closed-form derivative to obtain the critical points, greatly reducing the complexity of Newton’s method.
Finally, we remove duplicated critical points using spatial hashing.

CPE-MFA is designed for seamless integration with multithreading to enhance its efficiency \update{in span filtration and critical point extraction from spans.
In these two stages,} each span can be processed independently, which naturally aligns with multithreading.

\subsection{Span Filtration}
\label{sec:span-filtration}
By applying the strong convex hull property of B-splines (c.f. \cref{sec:MFA}), we are able to quickly determine that some regions in the domain cannot possibly contain a critical point. This allows us to exclude these (potentially large) regions from our fine-grained search for critical points.

First, we note that the derivative of B-spline function is also a B-spline function~\cite{piegl1997nurbs}. 
The partial derivative of \cref{eq:B_spline_2d} with respect to the first dimension is:
\begin{equation}
\label{eq:b-spline-derivative}
    \begin{aligned}
    &\frac{\partial}{\partial \modify{u_1}}F(u_1,\modify{\cdots},u_d) = \\
    &\sum_{j_1=0}^{n_1-2}\sum_{j_2=0}^{n_2-1}\cdots\sum_{j_d=0}^{n_d-1} N^{(1)}_{j_1, p-1}(u_1) N^{(2)}_{j_2, p}(u_2)\cdots N^{(d)}_{j_l, p}(u_d) \tilde{P}_{j_1,\cdots,j_d},
    \end{aligned}
\end{equation}
where
\begin{equation}
    \tilde{P}_{j_1,\modify{\cdots},j_d} = \frac{p}{t^{(1)}_{j_1+p+1} - t^{(1)}_{j_1+1}} \left(P_{j_1+1,j_2,\modify{\cdots}, j_d} - P_{j_1,\modify{\cdots},j_d}\right).
\end{equation}
Partial derivatives in the other directions follow by analogy. We abbreviate the derivative notation as $\partial_l F = \frac{\partial}{\partial \modify{u_l}}F$.

To locate all critical points of the B-spline defined in~\cref{eq:MFA}, we will identify points where every first partial derivative is zero. But since the partial derivatives of a B-spline are also B-splines, the problem of finding vanishing partial derivatives simply reduces to the problem of finding roots of the  associated ``derivative B-spline.''

For each dimension, we compute the derivative spline using~\cref{eq:b-spline-derivative}. Next, we scan the control points of the derivative spline to find regions in the domain that cannot contain zero values; if the derivative spline is not zero in this region, then there cannot be a critical point in this region. To achieve this, we utilize the strong convex hull property.

Consider an individual knot span and a direction $l$. From \cref{theorem:convex-hull}, we know that throughout this entire span, the value of the derivative spline $\partial_l F$ is within the convex hull of its neighboring control points. Should this convex hull fail to encompass zero, it must follow that $\partial_l F$ never vanishes within this knot span. Thus, this knot span cannot contain a critical point.
Therefore, by iterating through each span and computing the convex hull of the associated control points, we can find spans guaranteed  to not harbor a zero derivative in at least one dimension.
By excluding all of these spans, we can focus our search only on spans that may contain critical points.

This span filtration stage significantly streamlines the computational effort by discarding spans that do not contain critical points. 
As a result, this focused strategy not only saves computational resources but also time, enhancing the efficiency of extracting critical points.

\subsection{Critical Point Extraction in a Single Span}
\label{sec:cp-extraction}

For a point $(u_1,\cdots, u_d)$ in the span $[u_{i_1},u_{i_1+1}]\times \cdots\times[u_{i_d},u_{i_d+1}]$, the polynomial function $f:=F(u_1,\cdots, u_d)$ is decided by control points $P_{h_1,\cdots, h_d}$. 
The derivative of this polynomial function $f$, described in Eq.~\eqref{eq:b-spline-derivative}, is crucial for extracting critical points in the span. 

Specifically, the critical points are identified by locating all points in the span whose first derivatives are zero in every dimension. 
\update{MFA provides accurate high-order derivatives. Given the availability of the first and second derivatives of $f$}, Newton's method becomes a natural choice \update{for finding zeros of the first derivative of MFA}.

Critical points within each span are computed using Newton's method, detailed in \cref{algorithm:newton}, with a set of initial points $X$ uniformly sampled in the span. \update{We sample $(p+1)^d$ initial points in each span.}
Let $i_{max}$ be an upper bound on the number of iterations using the Newton's method. 
\modify{We set $i_{max}=20$ to ensure convergence, although the average number of iterations is often much lower as shown in \cref{tab:epsilon}.}
During an iteration of Newton's method, if a point moves beyond a certain distance $\xi$ from the center of the span, then it is unlikely to return (as a critical point), prompting the termination of the iteration to enhance the efficiency of the algorithm. 
 \modify{In our experiments, setting $\xi$ to be five times the span's diagonal length appears to be adequate}. 
We focus on extracting non-degenerate, isolated critical points with non-zero ($\geq \delta$) determinant of the Hessian matrix. 
\modify{We choose $\delta = 10^{-13}$ to be a sufficiently small positive number that strikes a balance between precision and efficiency.}

Throughout the process, when we identify a critical point within a span, we compare it against the set of already identified critical points $Z$ within the span. 
If the distance from newly identified critical point $x$ to the existing points in $Z$ is less than $\tau$, we consider this point to be a duplicate and ignore it. 
Otherwise, if $x$ is within the span, we include it into the list of critical points $Z$. \update{ $\varepsilon$ is the threshold to decide if the gradient is small enough to stop the iteration. $\tau$ and $\varepsilon$ are user-defined parameters discussed in \cref{sec:results}.}

\vspace{-2mm}
\begin{algorithm}[h]
    \caption{Critical Point Extraction in a Single Span}
    \label{algorithm:newton}
    \begin{algorithmic}[1]
    \REQUIRE 
    An input span $s$, and a function $f$ defined on the span.
    \ENSURE
    A set of critical points $Z$ in the span $s$.
    \STATE $Z = \emptyset.$
    \STATE Uniformly sample a set of initial points $X$ in the span $s$.
    \STATE Compute the center $c$ of the span $s$.
    \FORALL{initial point $x \in X$}
    \STATE $i=0$, $x_i = x$
    \WHILE{$i < i_{max}$} 
    \IF{$\mathrm{det}(H(x_i))<\delta$}
    \STATE \textbf{break}
    \ENDIF
    \STATE $x_{i+1}=x_i-H^{-1}(x_i) \nabla f(x_i)$, $H$ is the Hessian matrix.
    \IF{$||x_{i+1}-c||>\xi$ }
    \STATE \textbf{break}
    \ENDIF
    \IF{$||\nabla f(x_{i+1})||<\varepsilon$}
    \IF{$||x_{i+1}-y|| \geq \tau, \forall y\in Z$ and $x_{i+1} \in s$}
        \STATE $Z = Z \bigcup x_{i+1}$
    \ENDIF
    \STATE \textbf{break}
    \ENDIF
    \ENDWHILE
    \ENDFOR
    \end{algorithmic}
\end{algorithm}  
    \vspace{-3mm}
\subsection{Duplication Removal}
\label{sec:duplication-removal}

After removing duplicated critical points within each span in \cref{algorithm:newton}, there may still be duplicates near the junctions of spans. 
We address this issue by introducing spatial hashing. 
Assuming that critical points within a distance less than $\tau$ are considered to be duplicates, our duplication removal algorithm ensures that such points are assigned to the same hash buckets. 
 
For a point $x=(x_1,\cdots,x_d)$, we would like to ``snap'' it to the nearest integer grid point and apply spatial hashing. 
To do so, let $k = \frac{x}{20\tau} = (k_1, \cdots, k_d)$. 
We consider a hash index associated with $x$, $\mathrm{I}=(\Ind_1,\cdots,\Ind_d)$.  
Depending on the approximation, each $\Ind_{i}$ can have two distinct values $\floor{k_i }$ and $\ceil{k_i}$. 
 We choose $k=\frac{x}{20\tau}$, and two rounding methods (floor and ceiling) so that $x$ shares at least one hash index with all points within a distance $\tau$. 

By choosing floor or ceiling for each dimension $I_i$, we obtain a set of hash indices, $\mathcal{I} = \{I\}$, where there are $2^d$ distinct hash indices in $\mathcal{I}$. 
For each $\mathrm{I} \in \mathcal{I}$, we employ \verb|boost::hash_combine()|  from the \emph{Boost C++ Libraries} as the hash function to compute the hash values $V$ for $\mathrm{I}$. 
Finally, we compare $x$ to all the points $y$ already in the hash buckets determined by $V$. 
If $x$ and $y$ are less than $\tau$ apart, then $x$ is flagged as a duplicate.  
The pseudocode for spatial hashing is available in the supplement.

\subsection{Time Complexity}
We give a summary of complexity here; see the supplement for details.  
In span filtration, $p(p+1)^{d-1}$ control points are utilized in every span of the first derivative. The cost of span filtration is $\mathcal{O}(dp^dn)$, where $n$ is the number of spans. Each iteration of Newton's method involves computing the Hessian and gradient at $\mathcal{O}(d^2p^d)$ and solving linear system at $\mathcal{O}(d^3)$. Since $(p+1)^d$ initial points are used in each span, the cost of finding critical points in all spans is $\mathcal{O}(i_{max}d^2p^{2d}n)$, whereas the spatial hashing takes $\mathcal{O}((2p)^ddn)$. When $p\geq 2$, the overall time complexity is $\mathcal{O}(i_{max}d^2p^{2d}n)$.
In practice, $p, d \ll n$ (and can be treated as constants).  
In all the experiments conducted, more than 95\% of the time was spent finding critical points using Newton's method, corroborating the analysis of time complexity.

\section{Experimental Results}
\label{sec:results}

We perform a sanity check for the validity of our framework with a synthetic dataset (\cref{sec:Schwefel}) followed by a number of scientific datasets in 2D (\cref{sec:cesm} and \cref{sec:s3d}) and 3D (\cref{sec:qmc} and \cref{sec:rti}) with complex topological features. 
We employ adaptive fitting MFA~\cite{peterka2018foundations} and use degree-$3$ and degree-$2$ polynomials for the synthetic and scientific datasets, respectively. After fitting an MFA model into a discrete dataset, the MFA model serves as the input and basis for all further experiments.

\para{Implementation.}
We conduct experiments on a desktop equipped with an Intel 3.5 GHz Core i9 CPU featuring 8 hardware cores and 8 threads, along with 32 GB of DDR4 RAM. Our code is compiled using g++ version 11.4.0 with -O3 optimization. 
We use threading building blocks (TBB) for parallelization. \update{TBB balances the workload in every thread dynamically.} The number of threads is set to be the number of hardware cores. 

\para{TTK.} 
The Topology ToolKit (TTK) \cite{tierny2018topology} is a toolbox for topological data analysis. 
TTK assumes that the input data is a piecewise linear (PL) scalar field $f$ defined on a PL manifold $M$ of dimension 2 or 3 (e.g., a mesh interpolating a discrete set of points).
$f$ has value at the vertices of $M$, and is linearly interpolated on the higher-order mesh elements.
In the PL setting, a vertex is \emph{regular} if both its lower link and upper link are simply connected; otherwise, it is a \emph{critical point} of $f$.  
TTK extracts critical points from $M$ that comply with a discrete gradient \cite{tierny2018topology}. 
We call these critical points the \emph{PL critical points}, to differentiate them from those extracted from MFA using our approach.  
From an implementation perspective, we call the $\verb|ttkScalarFieldCriticalPoints()|$ function from TTK \cite{vidal2021progressive} to extract PL critical points from a PL function defined on a mesh. 

\para{Evaluation.} 
Our initial objective is to evaluate whether our CPE-MFA framework could extract all critical points from each scientific dataset. 
However, since the ``ground truth'' critical points from each dataset are unknown, we evaluate the output of CPE-MFA against those obtained by applying TTK to an MFA model. 
We aim to demonstrate that critical points extracted by our approach align well with the expectation. 
That is, the critical points extracted from an MFA model using our approach are similar to those extracted by a different critical point extraction method (e.g., TTK) applied to the same MFA model. \modify{The results of TTK are used as a reference, not as the ``ground truth''.}

First, we apply MFA to obtain a continuous implicit model. 
Second, we extract critical points from the MFA model using our approach based on Newton's method; this process is called CPE-MFA.  
Third, since TTK is not natively equipped to extract critical points from continuous implicit models, we construct a PL dataset by sampling a set of discrete points from the MFA model, and apply TTK to compute its PL critical points. \update{We assume that a higher resolution PL dataset aligns more with a continuous model and captures the critical points more accurately.} This process is called TTK-MFA. 
In summary, CPE-MFA processes critical points as elements of a continuous model, whereas TTK-MFA operates on a PL dataset sampled from an MFA model. 

Despite the fundamental difference between these two approaches, the TTK-MFA output provides a valuable reference for assessing the effectiveness of CPE-MFA. 
We demonstrate across all scientific datasets in this paper that despite minor differences, CPE-MFA critical points are similar to those from TTK-MFA.

\para{Technical details.}
For TTK-MFA, we sample a PL dataset (with a grid size of 1) from the MFA model that is identical in size to the original input data, to maintain consistency in our comparative analysis. 
\update{Thus, in \cref{sec:cesm,sec:s3d,sec:qmc,sec:rti},} we set a threshold $\tau=0.999$ for duplication removal in CPE-MFA, \update{based on the grid size 1 to sample the PL dataset for TTK-MFA}.  
TTK-MFA may identify some points on the boundary of the domain to be critical points (referred to as \emph{boundary critical points}) due to partial neighborhood information. 
However, it cannot be ascertained that these points are genuine critical points with zero gradient. 
Therefore, we exclude boundary critical points identified by TTK-MFA in our result presentation. 

\update{To quantitatively evaluate the similarity, we utilize the Jaccard index, defined as $\frac{|A\cap B|}{|A|+|B|-|A\cap B|}$, where $|A|$ and $|B|$ are the number of critical points from CPE-MFA and TTK-MFA, respectively, and $|A\cap B|$ denotes the number of critical point alignments.  A critical point from CPE-MFA is aligned with one from TTK-MFA if they are less than a grid size apart from each other (and thus considered to be co-located).}

\begin{figure}[!ht]
     \centering
     \includegraphics[width=\linewidth]{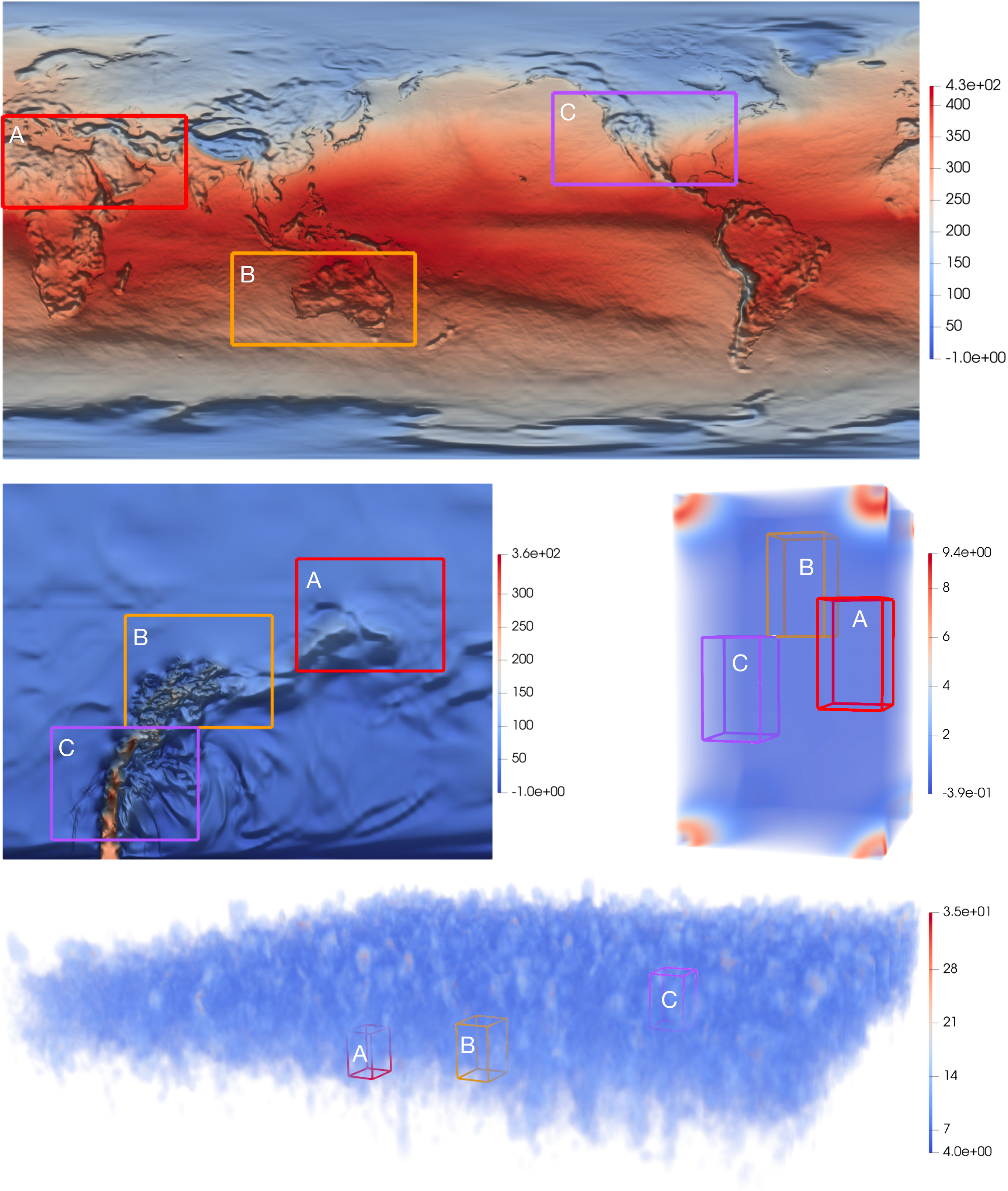}
     \vspace{-6mm}
     \caption{Scientific datasets: CESM (top), S3D (middle left), QMC (middle right), and RTI  (bottom).}
     \label{fig:scientific-data}
     \vspace{-4mm}
 \end{figure}

\para{Datasets and blocks.}
For each scientific dataset, visualized in \cref{fig:scientific-data}, we extract three blocks of data to highlight the extracted critical points from both CPE-MFA and TTK-MFA. \modify{The sizes and locations of these blocks are described in the supplement.}

\begin{table}[!ht]
\scriptsize
\centering
\vspace{-2mm}
\begin{tabu}{*{4}{c}}
\toprule
Dataset & \CellWithForceBreak{Original \\ \#Spans} &\CellWithForceBreak{Actually Evaluated \\\#Spans (\%)} &\CellWithForceBreak{Skipped \\\#Spans (\%)} \\
\midrule
CESM & 276740 & 43039 (15.55\%) & 233701 (84.45\%)  \\
S3D & 44377 & 9908 (22.33\%) & 34469 (77.67\%) \\
QMC & 18432 & 8229 (44.65\%) & 10203 (55.35\%)\\
RTI & 1741932 & 1707319 (98.01\%) & 34613 (1.99\%)\\
\bottomrule
\end{tabu}
\vspace{-2mm}
\caption{Results of span filtration, reporting the number of original spans, the number of spans (and percentage) actually being evaluated, and the number of spans (and percentage) being skipped during the optimization.}
\label{tab:filtered-span}
\vspace{-2mm}
\end{table}

\para{Span optimization.}
We report the number of spans identified in each dataset before and after the span filtration process, as shown in \cref{tab:filtered-span}.
Our optimization process ignores a substantial number of spans (up to 84\%).  
Even for the most complex RTI dataset (\cref{sec:rti}), a number of spans are excluded from the final phase of critical point extraction.
These exclusions highlight the effectiveness of our span filtration process in reducing irrelevant or redundant spans, thus substantially accelerating the overall critical point extraction algorithm.

\para{Running time.} 
Finally, we report the running time of CPE-MFA and TTK-MFA for all scientific datasets in \cref{tab:time}.

\begin{table}[!ht]
  \scriptsize
  \centering
  \resizebox{\columnwidth}{!}{
  \begin{tabu}{c|cc|cc}
  \toprule
  Block & \CellWithForceBreak{CPE-MFA\\ original} & \CellWithForceBreak{TTK-MFA\\ original} & \CellWithForceBreak{CPE-MFA\\ upsample} &\CellWithForceBreak{TTK-MFA \\ upsample} \\
  \midrule
  \multicolumn{5}{c}{CESM}\\
  \midrule
  Entire domain &  3.30 & 1.66 & 3.43 & -\\
  A& 0.314 & 0.0759 & 0.391 & 7.305 \\
  B & 0.187 & 0.0724 & 0.189 & 7.720 \\
  C & 0.122 & 0.0823 & 0.122 & 7.306\\
  \midrule
  \multicolumn{5}{c}{S3D}\\
  \midrule
  Entire domain & 0.554 & 0.101 & 0.560 & -\\
  A & 0.017 & 0.0256 & 0.017 & 0.756 \\
  B & 0.169 & 0.0228 & 0.169 & 0.772\\
  C & 0.169 & 0.0289 & 0.167 & 0.773\\
  \midrule
  \multicolumn{5}{c}{QMC}\\
  \midrule
  Entire domain & 3.310 & 0.517 & 3.375 & -\\
  A & 0.180 & 0.0316 & 0.222 & 15.247\\
  B & 0.149 & 0.0304 & 0.151 & 15.646\\
  C & 0.159 & 0.0316 & 0.161 & 16.002\\
   \midrule
  \multicolumn{5}{c}{RTI}\\
  \midrule
  Entire domain & 1003.1 & 15.50 & 985.1  & -\\
  A & 0.887 & 0.0186 & 0.876 & 1.823 \\
  B & 1.079 & 0.0165 & 1.124 & 1.840 \\
  C & 1.209 & 0.0189 & 1.165 & 1.973\\
  \bottomrule
  \end{tabu}
  }
  \vspace{-2mm}
  \caption{Running time (in seconds) across all scientific datasets. The upsampling ratio is at $10^2$ for the 2D and $10^3$ for the 3D datasets.}
  \label{tab:time}
  \vspace{-4mm}
  \end{table}
  
\subsection{Schwefel Dataset}
\label{sec:Schwefel}
A (scaled) Schwefel function is a non-convex function that can be generated in any dimension \cite{schwefel1981numerical}, that is, for $\mathbf{x}=[x_1, x_2, \dots, x_d]$, 
\begin{equation}
\label{eq:schwefel}
    f(\mathbf{x})=\frac{1}{2}\left(418.9829 d -\sum_{i=1}^d x_i\sin(\sqrt{|x_i|})\right),
\end{equation}
where $d$ is the dimension. 
It serves as a good synthetic dataset as it is complex and contains many local minima. 
There are $(2k-2)^d$ non-degenerate critical points in the domain $[-\left((k+\frac{1}{2})\pi\right)^2, \left((k+\frac{1}{2})\pi\right)^2]^d$, where $k$ is a natural number. We scale the function by $1/2$ for a better visualization. \update{Among all the critical points, there are $(k-1)^d$ local maxima, $(k-1)^d$ local minima, and $2(k-1)^d$ saddles.}

\begin{figure}[!ht]
    \centering
    \vspace{-4mm}
    \includegraphics[width=0.8\linewidth]{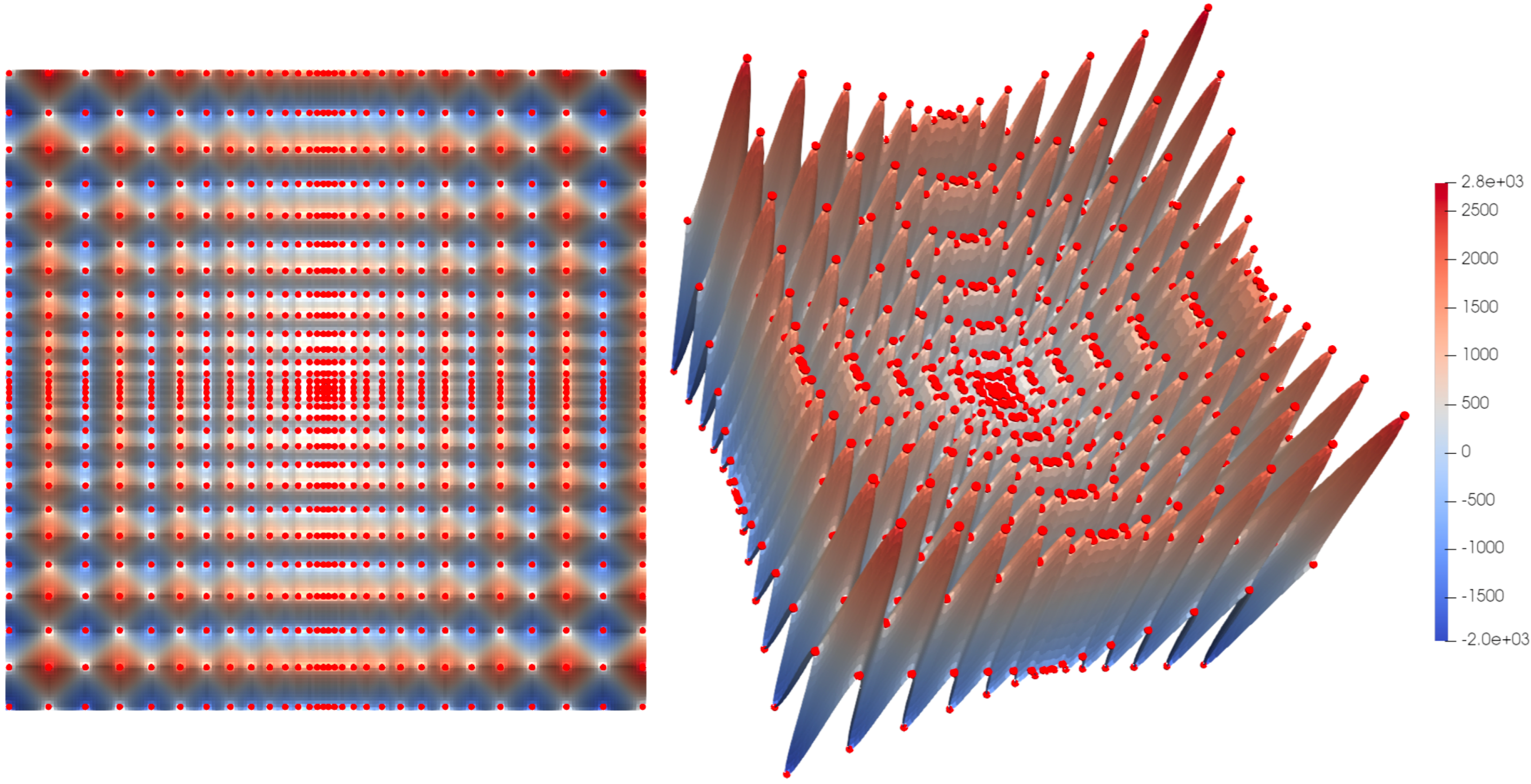}
    \vspace{-4mm}
    \caption{The Schwefel function with red critical points identified by CPE-MFA: (left) top view (right) side view.}
    \label{fig:schwefel}
    \vspace{-2mm}
\end{figure}

\begin{figure*}[!ht]
    \centering
    \vspace{-0.5mm}
    \includegraphics[width=\linewidth]{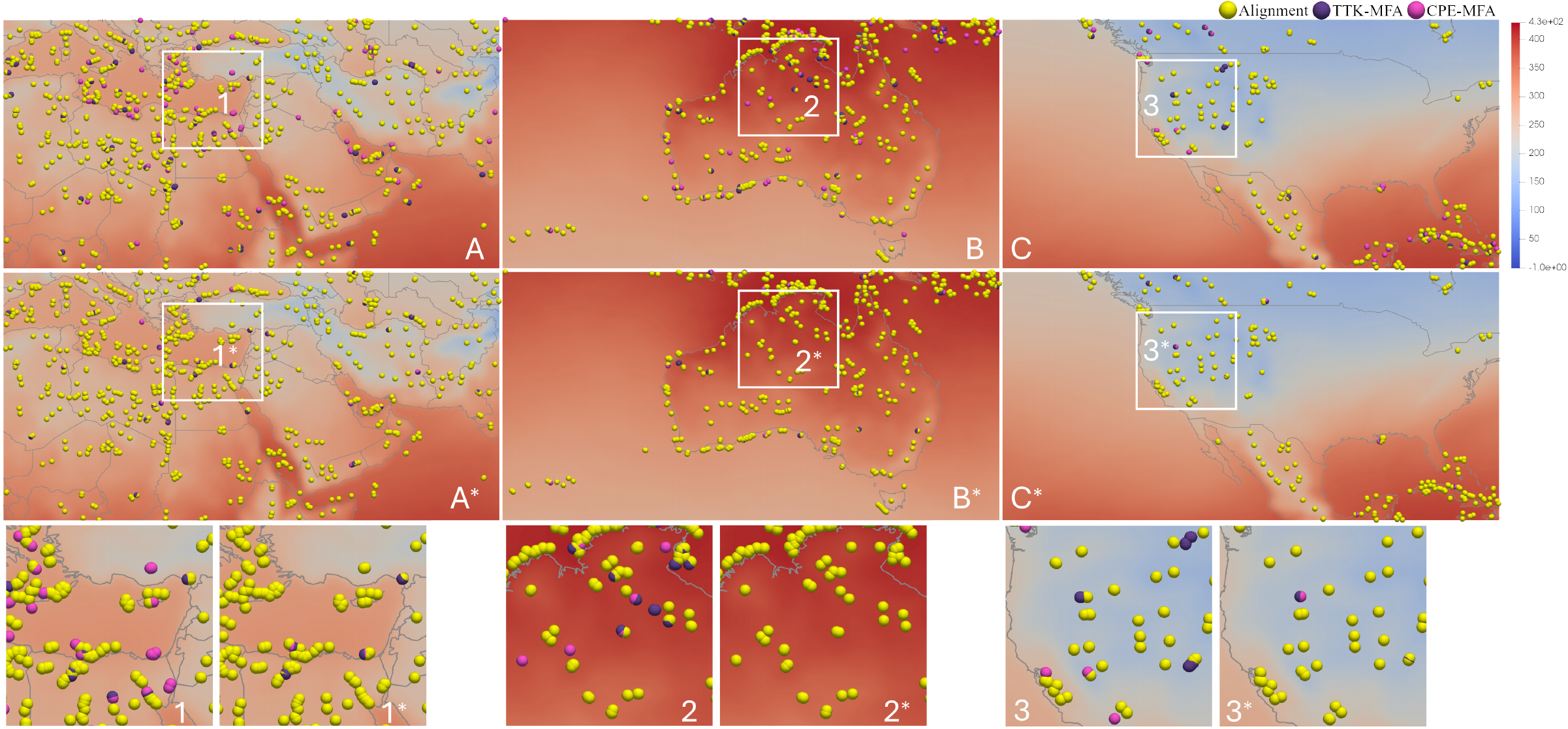}
    \vspace{-6mm}
    \caption{CESM dataset with critical points identified by CPE-MFA and TTK-MFA. Top: critical points from blocks A, B, and C, respectively. Middle: critical points from upsampled blocks A, B, and C (labeled as A$^*$, B$^*$, and C$^*$), respectively, with a ratio of $10^2$. Bottom: zoomed-in views of regions in the domain with and without upsampling (at a ratio of $10^2$).}
    \label{fig:cesm}
    \vspace{-2mm}
\end{figure*}

\begin{figure*}[!ht]
    \centering
    \includegraphics[width=0.9\linewidth]{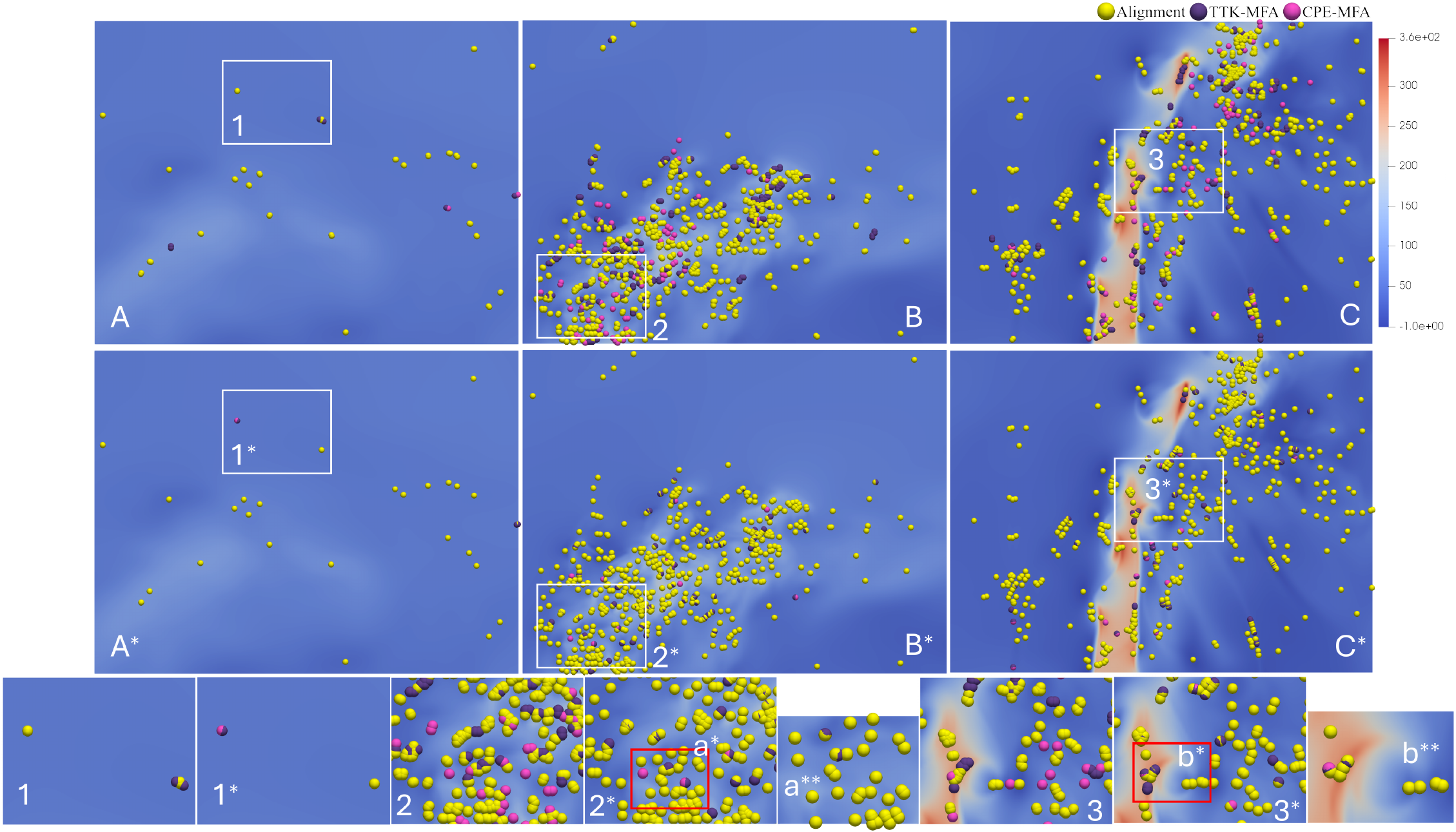}
    \vspace{-2mm}
    \caption{S3D dataset with critical points identified by CPE-MFA and TTK-MFA. Top: critical points from blocks A, B, and C, respectively. Middle: critical points from upsampled blocks A, B, and C (labeled as A$^*$, B$^*$, and C$^*$), respectively,  with a ratio of $10^2$. \update{Bottom: ($1, 1^*, 2, 2^*, 3,3^*$):  zoomed-in views of regions in the domain with and without upsampling at a ratio of $10^2$; (a$^{**}$, b$^{**}$): better critical point alignments (in regions a$^{*}$, b$^{*}$, respectively) after upsampling at a ratio of $10^4$.}}
    \label{fig:s3d}
    \vspace{-4mm}
\end{figure*}

To evaluate the efficacy of our method, we generate a \emph{Schwefel dataset} by sampling from a 2D Schwefel function uniformly, selecting $200^2$ points within the domain $\left[-2400,2400\right]^2$. 
Theoretical analysis predicts the existence of $900$ critical points in this domain. 
We apply MFA to this dataset with a polynomial degree of $3$ and $100^2$ control points.  
Using a span filtration, \update{$2809$ spans are identified for critical point extraction out of all $9409$ spans.} 
As shown in \cref{fig:schwefel}, CPE-MFA accurately identifies $900$ critical points, aligning perfectly with the theoretical expectation. \update{Meanwhile, CPE-MFA successfully classifies them into 225 local minima, 225 local maxima, and 450 saddles. We set $\tau$ to be $10^{-4}$ times the minimum width of each dimension in the domain.}

\update{Utilizing this dataset, in \cref{tab:epsilon}, we illustrate the impact of $\varepsilon$ on both the number of iterations (required by Newton's method) and the accuracy of the extracted critical points, accessed by the magnitude of the gradient. 
The theoretical average gradient magnitude of critical points is zero. 
Setting $\varepsilon=10^{-7}$ achieves an optimal balance between the number of iterations and the accuracy of the extracted critical points. This $\varepsilon$ is tested across several synthetic datasets and consistently performs well. Therefore, we use $\varepsilon=10^{-7}$ for all the experiments.}

\begin{table}[!ht]
\scriptsize
\centering
\begin{tabu}{*{6}{c}}
\toprule
$\varepsilon$ & $10^{-3}$&$10^{-5}$ & $10^{-7}$ & $10^{-9}$ & $10^{-11}$\\
\midrule
\CellWithForceBreak{Avg \#itr} & 3.16 & 3.51 & 3.78 & 3.91 & 19.34\\
\CellWithForceBreak{Avg grad mag}&$9.00e^{-5}$&$7.91e^{-7}$&$2.20e^{-9}$&$1.36e^{-10}$&-\\
\bottomrule
\end{tabu}
\vspace{-2mm}
\caption{\update{Average number of iterations and average gradient magnitude of the extracted critical points with different $\varepsilon$. When $\varepsilon=10^{-11}$, Newton's method cannot find all critical points within $i_{max}=20$ iterations.}}
\label{tab:epsilon}
\vspace{-6mm}
\end{table}

\begin{table*}[!ht]
    \scriptsize
    \centering
    \begin{tabu}{c|cccc|cccc}
    \toprule
        Block & 
        \CellWithForceBreak{CPE-MFA\\ original \#cp} & \CellWithForceBreak{TTK-MFA\\ original \#cp} & \CellWithForceBreak{Aligned \\ \#cp / type} & 
         \CellWithForceBreak{Jaccard index\\ original} & 	
         \CellWithForceBreak{CPE-MFA\\ upsample \#cp} &
         \CellWithForceBreak{TTK-MFA \\ upsample \#cp} & \CellWithForceBreak{Aligned \\ \#cp / type} &
         \CellWithForceBreak{Jaccard index\\ upsample}\\ 
    \midrule
    \multicolumn{9}{c}{CESM (2D)}\\
    \midrule
    Entire domain & 7940 & 8759 & 7204 / 7149 &0.76& 8549 & - & -&-\\
    A & 714 & 765 & 641 / 636 & 0.76 & 783 & 865 & 767 / 766 & 0.87\\
    B & 421 & 461 & 374 / 373 & 0.74 & 473 & 527 & 461 / 461 & 0.86\\
    C & 208 & 218 & 183 / 182 & 0.75 & 228 & 254 & 223 / 222 & 0.86\\
    D & 623 & 676 & 553 / 551  & 0.74 & 688 & 772 & 664 / 661 & 0.83\\
    E & 636 & 703 & 591 / 588 & 0.79 & 684 & 801 & 674 / 672 & 0.83\\
    \midrule
    \multicolumn{9}{c}{S3D (2D)}\\
    \midrule
    Entire domain & 1317 & 1732 & 1150 / 1116 & 0.61& 1569 & - & - & -\\
    A & 27 & 31 & 25 / 24 &0.76& 27 & 36 & 26 / 26 & 0.70\\
    B & 485 & 646 & 415 / 405 & 0.58& 608 & 813 & 596 / 595 & 0.72\\
    C & 450 & 552 & 379 / 363 & 0.61& 562 & 717 & 541 / 536 & 0.73\\
    D & 72 & 98 & 70 / 68 & 0.70 & 74 & 100 & 73 / 73 & 0.72 \\
    E & 44 & 62 & 43 / 43 & 0.68 & 45 & 63 & 44 / 44 & 0.69 \\
    \midrule
    \multicolumn{9}{c}{QMC (3D)}\\
    \midrule
    Entire domain & 210 & 262 & 167 / 161 &0.55& 244 & - & -&-\\
    A & 64 & 61 & 46 / 45 &0.58& 75 & 91 & 73 / 73 & 0.78\\
    B &  14 & 20 &  13 / 11 &0.62& 14 &  52 & 14 / 14 & 0.27\\
    C & 23 & 20 & 17 / 17 &0.65& 26 & 29 & 25 / 25 & 0.83\\
    D & 9 & 17 & 9 / 9 &0.53& 9 & 13 & 8 / 8 & 0.57\\
    E & 9 & 10 & 7 / 5 &0.58& 10 & 10 & 10 / 10 & 1.00\\
    \midrule
    \multicolumn{9}{c}{RTI (3D)}\\
    \midrule
    Entire domain & 711504 & 670156 & 425906 / 292248 & 0.45& 1258055 & - & -&-\\
    A & 118 & 81 & 61 / 50 & 0.44 & 181 & 253 & 161 / 156 & 0.59\\
    B & 438 & 292 & 256 / 172 &0.54& 827 & 1352 & 735 / 710 & 0.51\\
    C & 415 & 344 & 280 / 185 &0.58& 807 & 1198 & 733 / 712 & 0.58\\
    D & 503 & 315 & 289 / 186 &0.55& 952 & 1482 & 858 / 825 &0.54\\
    E & 467 & 346 & 300 / 211 &0.58& 852 & 1223 & 771 / 753 &0.59\\
    \bottomrule
    \end{tabu}
    \vspace{-2mm}
    \caption{Number of critical points (\#cp) extracted by CPE-MFA and TTK-MFA within each of the four scientific datasets as well as their corresponding selected blocks: original \#cp shows the number of critical points at the original resolution; upsample \#cp shows the number of critical points at an upsampling ratio of $10^2$ for the 2D and $10^3$ for the 3D datasets; aligned \#cp shows the number of critical points perfectly aligned between the two methods; \modify{aligned type shows the number of aligned critical points that share the same critical point type}; Jaccard index reports the similarity between the two sets of critical points extracted from CPE-MFA and TTK-MFA respectively. Critical points from TTK-MFA in the upsampling setting are not reported as it runs out of memory when processing the large data. We add two additional randomly selected blocks (block D and block E) per dataset to collect more statistics on the distributions of critical points.}
    \label{tab:cpt}
    \vspace{-4mm}
    \end{table*}

\subsection{CESM Dataset}
\label{sec:cesm}

The Community Earth System Model (CESM) offers comprehensive global climate data, spanning the Earth’s land, oceans, atmosphere and sea ice. 
Our analysis focuses on the FLDSC variable, which represents the clear-sky downwelling long-wave flux on the surface, within the Community Atmosphere Model (CAM) developed at the National Center for Atmospheric Research (NCAR) \cite{neale2010description}. 
The CESM dataset we use comprises a $3600 \times1800$ 2D domain, with each grid point representing a singular value of the FLDSC variable.

For qualitative and quantitative analysis, \cref{fig:cesm} visualizes the critical points identified by both methods, whereas \cref{tab:cpt} reports their numbers, \modify{position and type  alignments}.
In \cref{fig:cesm,fig:s3d,fig:qmc,fig:rti}, critical points where CPE-MFA and TTK-MFA align are depicted in yellow.  The critical points exclusive to TTK-MFA are colored in purple, and those exclusive to CPE-MFA are shown in pink. In \cref{fig:cesm} A, B, and C, the extracted critical points from CPE-MFA and TTK-MFA mostly align with each other \update{(observing all yellow points),} with small discrepancies between the two methods across all three blocks. 
The Jaccard indices between these two sets of critical points are shown in \cref{tab:cpt} (original), valued at $0.76$ for the entire dataset, and $0.76, 0.74, 0.75$ for blocks A, B, and C, respectively. 

There are minor discrepancies (misalignments) in certain regions where critical points are detected by one but not the other method \update{(see independent purple or pink points within the boxes $1, 2,$ and $3$, with zoomed-in views).  }
We hypothesize that these discrepancies are due to the sampling process that generates a PL dataset for TTK, which may  not capture all the features from the continuous MFA model and \modify{may contain spurious critical points due to the PL approximation}.  
We hypothesize further that upsampling the PL dataset will improve the alignment between the two methods. 

To validate this hypothesize, we increase the sampling resolution for TTK to better preserve the features of MFA. 
By applying an upsampling ratio of $10^2$, we expand the size of the PL dataset in every dimension by a factor of 10 relative to the original sampling resolution. 
Critical points extracted using TTK from the upsampled PL dataset demonstrate improved alignment between the two methods, see \cref{fig:cesm} A$^*$, B$^*$, and C$^*$, as well as zoomed-in boxes: $1$ vs $1^*$, $2$ vs $2^*$, and $3$ vs $3^*$. 
As shown in \cref{tab:cpt} (upsample), the number of perfectly aligned critical points increases with upsampling, and the Jaccard indices (upsample) also increase from around $0.75$ to $0.86$ across all three blocks. \update{For points that do not align, we observe closely situated points in different colors.}

For the remaining misaligned critical points, an increased upsampling ratio of $10^4$ gives rise to perfectly aligned critical points between the two methods in these regions (results not shown). 

In summary, the improved alignment with upsampling validates our hypothesis regarding the initial discrepancies. 

\para{Details on critical point alignment.} 
\cref{tab:cpt} reports the number of critical points identified by CPE-MFA and TTK-MFA, respectively. 
The column TTK-MFA original lists the number of critical points identified by TTK-MFA in the original resolution $(u=1)$.
To ensure comparability between the results of CPE-MFA from a continuous model and those of TTK-MFA from the corresponding PL dataset, we align $\tau$ with the grid size.
Thus, the column CPE-MFA original displays the number of critical points detected by CPE-MFA at a duplicate removal threshold $\tau=0.999$.
The column TTK-MFA upsample reveals the number of critical points identified by TTK-MFA after upsampling at a ratio $u=10^2$. 
Correspondingly, the column CPE-MFA upsample uses a threshold of $\tau=0.0999$ under the same upsampling condition. 

\para{Highlighted results.} 
We would like to emphasize that we do not expect the critical points to align perfectly between the two methods, in both locations and quantities, since they are fundamentally different extraction methods.  
Using TTK-MFA extracted critical points as a reference, we demonstrate a strong alignment between the two sets of critical points (\modify{in terms of location and type}), indicating the validity of our CPE-MFA framework. 
Furthermore, as we increase the upsampling ratio for TTK, the alignment between the two sets of critical points improves, validating our hypothesis that the discrepancies are due to the sampling resolution (thus approximation quality)  for TTK. 

\subsection{Turbulent Combustion Dataset}
\label{sec:s3d}

The dataset is from a turbulent combustion S3D simulation \cite{chen2009terascale} that models the combustion of a fuel jet influenced by an external cross flow \cite{fric1994vortical,grout2011direct,peterka2022multivariate}. 
The simulation uses a 3D domain of size $704 \times 540 \times 550$. 
The variable of interest represents the magnitude of the 3D velocity within the domain. 
For our experiment, we use a 2D cross-section of the dataset of size $704 \times 540$.
 
The critical points extracted by CPE-MFA and TTK-MFA are visualized in \cref{fig:s3d} A, B, and C across all three blocks. 
Mirroring similar observations from the CESM dataset, we observe reasonable critical point alignments between the two methods, despite minor discrepancies; see the zoomed-in views of boxes \update{$1$, $2$, and $3$} with blocks A, B, and C, respectively. 
The number of critical points and their similarities (reported as Jaccard indices) are shown in \cref{tab:cpt}. 

Again, these discrepancies can be mitigated following an upsampling for TTK at a ratio $u=10^2$. 
\update{As shown in the zoomed-in boxes 2 vs. 2$^*$ and 3 vs. 3$^*$ in \cref{fig:s3d}, the critical points from both methods become more aligned in these regions with upsampling. In box 1$^*$, even though each method has an unaligned point, the points nearly overlap with each other. The distance between these two points is smaller than the distance between the corresponding points in box 1. Thus, points in box 1$^*$ is more aligned than in box 1.}
The Jaccard indices among the two sets of critical points also increase drastically with upsampling, as shown in \cref{tab:cpt}. 

A few discrepancies remain (at an upsampling resolution $10^2$), they are highlighted in regions enclosed by small red boxes labeled as $a^*$ and $b^*$. 
Critical points in these regions become more aligned with further upsampling at a ratio of $10^4$; see the zoomed-in views of a$^{**}$, b$^{**}$ in \cref{fig:s3d}. For all unaligned points, nearly overlapping points in different colors can be found. As the threshold for alignment decreases with upsampling, the criteria for alignment become increasingly stringent. For critical points in a$^{**}$ and b$^{**}$, even if they do not align, they remain very close to each other. 


\para{Highlighted results.} 
Recall that TTK-MFA extracts critical points from PL dataset sampled from an MFA model, whereas CPE-MFA extracts critical points directly from the same continuous MFA model. 
The output from TTK-MFA depends on the sampling resolution, whereas the output of CPE-MFA does not depend on discretization or sampling. 
However, we do observe an increase in the number of critical points from CPE-MFA as we increase the sampling resolution. This is because we decrease the threshold used to remove duplicates to be aligned with the upsampling resolution.  
The key observation is that as the sampling resolution increases, the result of TTK-MFA becomes more aligned with that of CPE-MFA. 
Since CPE-MFA operates directly on a continuous model, its output is not severely impacted by discretization or sampling, as does TTK-MFA.

\subsection{Quantum Monte Carlo Dataset}
\label{sec:qmc}

The QMC (or QMCPACK) dataset~\cite{zhao2020sdrbench} comes from an open-source Quantum Monte Carlo program designed for high-level \emph{ab initio} calculations of the electronic structure in atoms, molecules, solids, and 2D  nanomaterials~\cite{kim2018qmc}. 
The dataset we use comes from a $69\times 69\times 115\times 288$ Einspline dataset in QMC.
We choose the $10$th orbit among all the $288$  orbitals as the 3D scaler field of interest.  
 
\begin{figure}[!ht]
        \centering
        \includegraphics[width=0.8\linewidth]{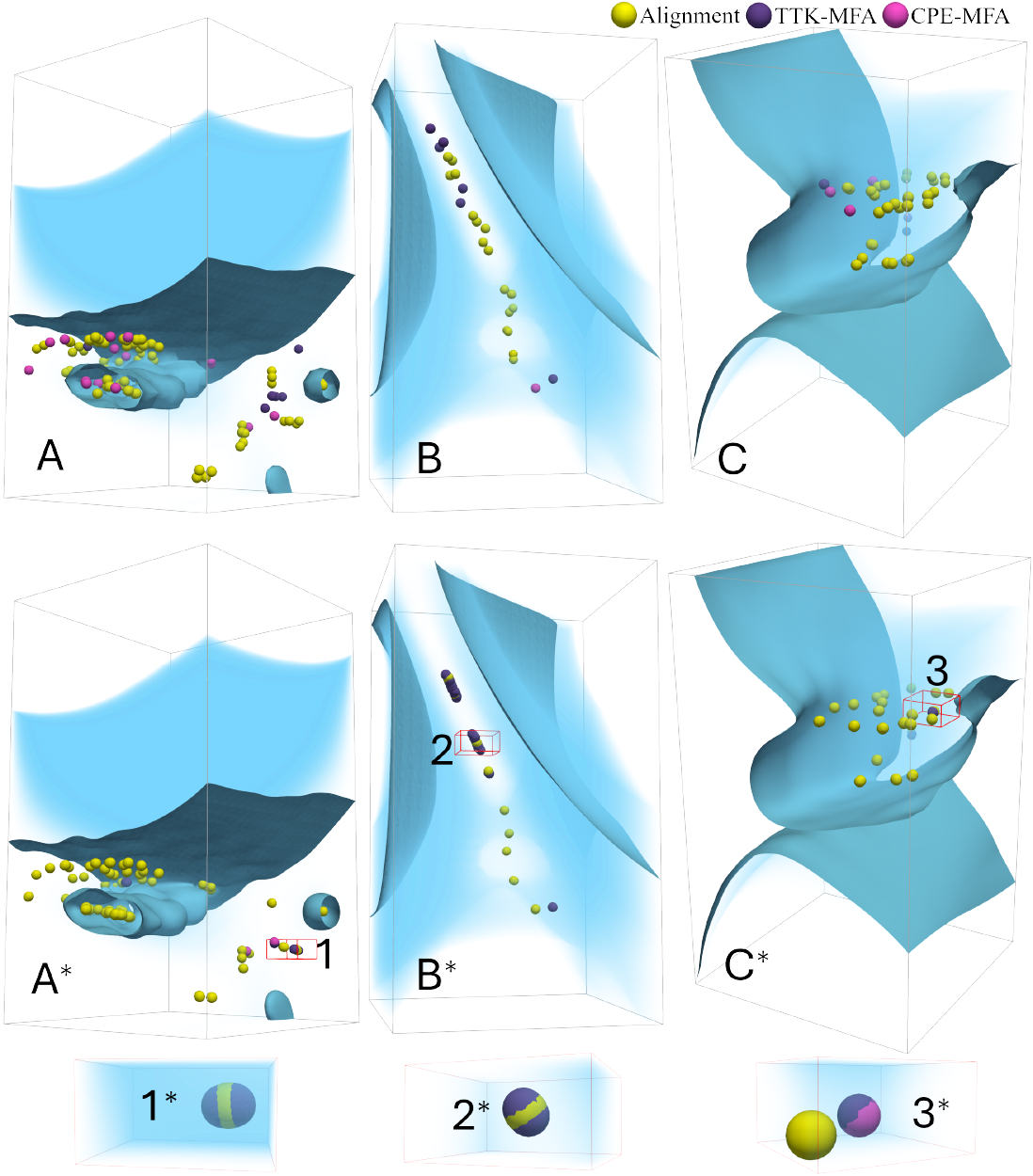}
        \vspace{-2mm}
        \caption{QMC dataset with critical points identified by CPE-MFA and TTK-MFA. \update{Top: critical points from blocks A, B, and C, respectively. Middle: critical points from upsampled blocks A, B, and C (labeled as A$^*$, B$^*$, and C$^*$), respectively, with a ratio of $10^3$. Bottom: better critical point alignments 1$^{*}$, 2$^{*}$, 3$^{*}$ (in regions 1, 2, 3, respectively) after upsampling at a ratio of $10^6$.}}
        \label{fig:qmc}
        \vspace{-4mm}
    \end{figure}

The results of this 3D dataset are presented in~\cref{fig:qmc} and \cref{tab:cpt}.
As shown in \cref{tab:cpt}, we observe that at the original resolution, there are some reasonable alignments between the two methods. Specifically, in the entire domain, 80\% of CPE-MFA critical points align with 64\% of TTK-MFA critical points, giving a Jaccard index of $0.55$ between the two sets of critical points. 

\update{To facilitate clearer observation, we display results only within specific value ranges. For block A in~\cref{fig:qmc}, points with value in $[-6.4e^{-6},1e^{-3}]$ are shown. For block B, the range is $[-2.5e^{-5},2e^{-3}]$, and for block C, it is $[-6.3e^{-6},6.3e^{-5}]$.
As shown in \cref{fig:qmc} A$^*$, B$^*$, and C$^*$ respectively, upsampling with ratio $10^3$ greatly reduces the discrepancies.}   

For an in-depth investigation, for the discrepancies observed with an upsampling ratio at $10^3$, we could apply a higher upsampling ratio at $10^6$ for TTK-MFA to obtain \update{further improved critical point alignment. 
For instance, within the red cube 1 from ~\cref{fig:qmc} A$^*$ at an upsampling ratio of $10^3$, there are unaligned points from both TTK-MFA and CPE-MFA. Upon increasing the sampling ratio to $10^6$, all points from CPE-MFA align, the unaligned TTK-MFA points are very close to an aligned CPE-MFA point, as shown in ~\cref{fig:qmc} 1$^{*}$.
Similar results appear in 2 vs. 2$^*$. In cube 3$^*$, corresponding to the red cube 3, the unaligned point from CPE-MFA and TTK-MFA are almost aligned.}

\para{Highlighted results.} 
The above results suggest that upsampling effectively bridges the gap between the sampled PL-dataset used by TTK-MFA and the continuous model used by CPE-MFA. 
This experiment confirms the efficacy of CPE-MFA in identifying critical points from MFA.

\subsection{Rayleigh-Taylor Instability Dataset}
\label{sec:rti}

When two fluids of different densities interact with each other, and the lighter fluid pushes the heavier fluid under constant acceleration, the phenomenon called the Rayleigh-Taylor instability \cite{livescu2010new} emerges. 
We use a 3D RTI dataset generated by simulating this instability with a CFDNS \cite{livescu2009cfdns} Navier-Stokes solver. 
We use the velocity vector magnitude as the scalar field of interest. 
We sample a single time step within a domain of size $144\times 256\times 256$.  

\begin{figure}[!ht]
    \centering
    \includegraphics[width=0.8\linewidth]{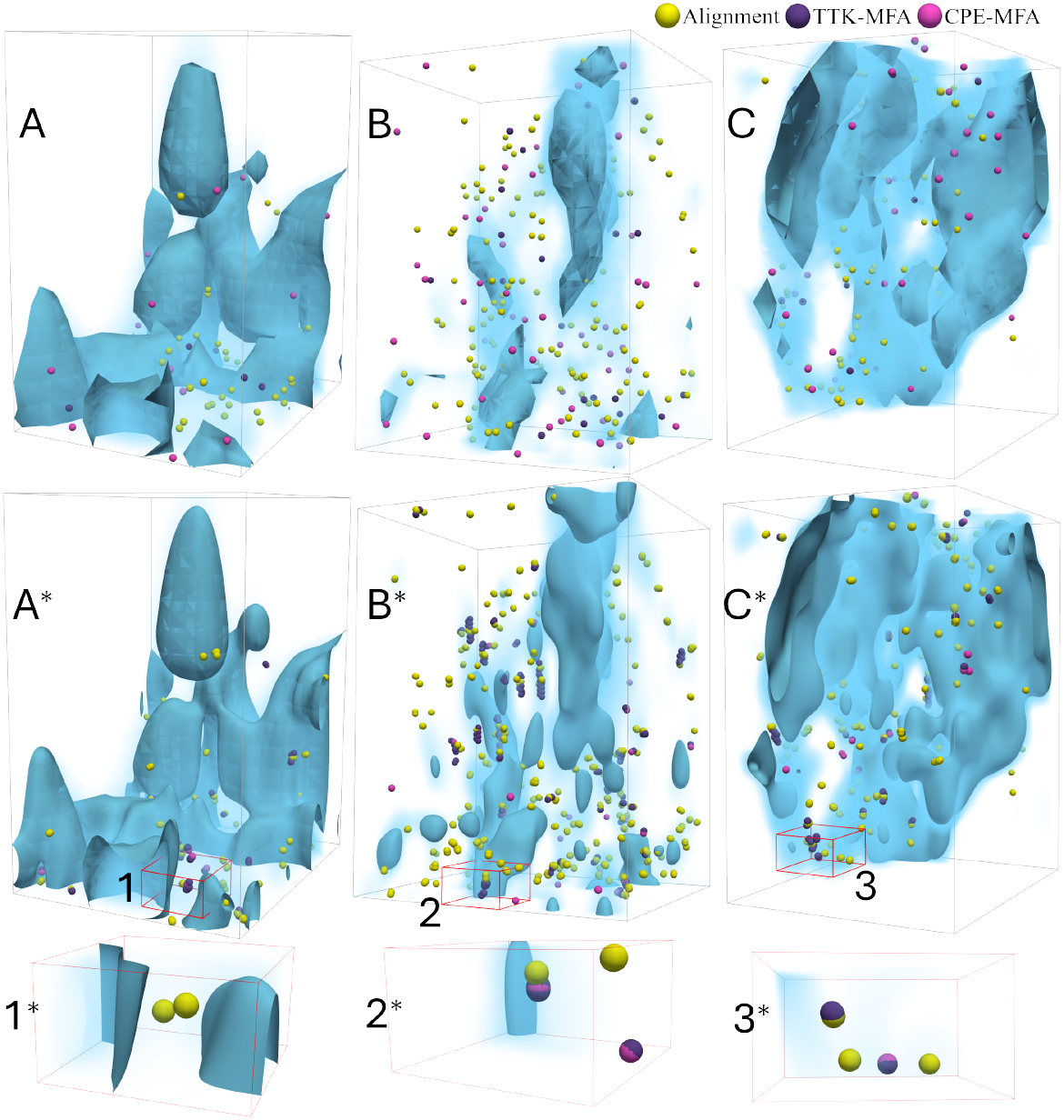}
    \vspace{-2mm}
    \caption{RTI dataset with critical points identified by CPE-MFA and TTK-MFA. \update{Top: critical points from blocks A, B, and C, respectively. Middle: critical points from upsampled blocks A, B, and C (labeled as A$^*$, B$^*$, and C$^*$), respectively, with a ratio of $10^3$. Bottom: better critical point alignments 1$^{*}$, 2$^{*}$, 3$^{*}$ (in regions 1, 2, 3, respectively) after upsampling at a ratio of $10^6$.}}
    \label{fig:rti}
    \vspace{-6mm}
\end{figure}

\cref{fig:rti} and \cref{tab:cpt} display the results of CPE-MFA and TTK-MFA across all three blocks.
This is a particularly challenging dataset due to the complexity of the flow, which gives rise to densely distributed critical points. 
\update{We set the display range for point value in blocks A, B, and C to $[8,16]$, $[2, 5]$, and $[8,13]$, respectively.}
We observe some critical point alignment, see \cref{fig:rti} A, B, and C. 
Upsampling with a ratio of $10^3$ has proven effective in reducing the discrepancies between the two methods; see \cref{fig:rti} A$^*$, B$^*$, and C$^*$.  

\update{Upsampling further at a ratio of $10^6$ aligns the critical points better in highlighted regions (red cubes 1, 2, and 3). 
All unaligned critical points in cube 1 become perfectly aligned  after upsampling, shown in 1$^{*}$. Compared with cube 2, the purple and pink points in 2$^*$ are almost overlapping. Although they do not align under the stringent threshold for the upsampled data, their positions are very close to one another. Similar results are observed in cubes 3 and 3$^*$.}

\para{Highlighted results.} 
The RTI dataset is topologically the most complex one in our experiments.  
We observe a reasonable alignment between the two sets of critical points extracted by CPE-MFA and TTK-MFA both qualitatively in \cref{fig:rti} and qualitatively in \cref{tab:cpt}, indicated by Jaccard indices between $0.44$ and $0.58$ in the original resolution.  
Upsampling for the TTK-MFA further improves the alignment, again, showcasing the efficacy of our method. 

\section{Conclusion and Discussion}
\label{sec:conclusion}

We introduce CPE-MFA, the first framework in extracting critical points from multivariate functional approximation (MFA) models of large-scale datasets. 
Our framework processes the continuous MFA models directly without resampling or discretization, utilizing multithreading. 
The framework's ability to bypass spans devoid of critical points further enhances its efficiency. 
Although our experiments focus on 2D and 3D datasets, our framework is dimension independent and generalizes easily to high-dimensional data. 
We demonstrate the effectiveness of CPE-MFA across various scientific datasets.  
This is the first step toward enabling continuous implicit models such as MFA to support topological data analysis at scale. 
In the future, we are interested in leveraging our current findings to extract topological descriptors from MFA models, such as contour trees and Morse complexes. 
\modify{We will also explore how critical points vary with different MFA approximation levels.}

Our framework has its limitations. 
\modify{Since TTK is a widely-used tool for critical point extraction, we compare our CPE-MFA results against those obtained by TTK based on a PL interpolation of points sampled from an MFA. 
Alternatively, we could also compare against trilinear interpolation (e.g., following Globus et al~\cite{globus1991tool}). 
Note that PL and  trilinear interpolation work with points sampled from an MFA model, thus they do not perfectly approximate the continuous domain. They serve as references, not as ground truth for critical point extraction.}
\modify{For certain dataset, even though the Jaccard index is low between CPE-MFA and TTK-MFA, (almost) all critical points from CPE-MFA are aligned with TTK-MFA. However, TTK-MFA tends to extract more (spurious) critical points than CPE-MFA (c.f.,~\cref{tab:cpt}) due to the noisy PL reconstruction of an MFA model (e.g., critical points lie in adjacent zigzags). Applying persistence simplification to the PL approximation might remove spurious critical points of TTK-MFA, thus increasing the Jaccard index. This is left for future work.}

Recall a critical point $x$ is \emph{isolated} if there is a neighborhood $U$ around $x$ and $x$ is the only critical point in $U$; otherwise, it is \emph{non-isolated}. Our framework only handles isolated critical points, but not non-isolated ones (e.g., a manifold of critical points). Additionally, we omit boundary critical points at the moment, which might not have zero derivatives. Furthermore, the choice of initial points for Newton's method may affect the results. For some MFA models, we may need more initial points in a span to extract all critical points.  
\modify{Finally, B\'ezier clipping~\cite{sederberg1990curve} (also known as interval Newton's method) offers an alternative approach for root finding with 
guaranteed convergence. It utilizes the convex hull properties of B\'ezier curves and iteratively subdivides the domain to narrow down the root-finding region. Replacing Newton's method with B\'ezier clipping in our framework will be an interesting future direction.}

\acknowledgments{
This work is supported by the U.S. Department of Energy (DOE), Office of Science, Office of Advanced Scientific Computing Research, under contract numbers DE-AC02-06CH11357, program manager Margaret Lentz. 
It is also supported in part by a grant from National Science Foundation (NSF) IIS-2145499.
}


\bibliographystyle{abbrv-doi}
\bibliography{refs-mfa}

\clearpage
\newpage
\appendix 
\section{Details on Scientific Datasets}
\label{sec:datasets}

For each scientific dataset, we extract three
blocks of data to highlight the extracted critical points from both CPE-MFA and TTK-MFA. The sizes and locations of these blocks
are detailed in \cref{tab:blocks}.

\begin{table}[ht]
  \scriptsize
  \centering
  \resizebox{1.0\columnwidth}{!}{
  \begin{tabu}{*{3}{c}}
  \toprule
  Dataset & Size & Block A\\
  \midrule
  CESM & $3600 \times 1800$ 
  & $[0,720] \times [989,1349]$ \\ 
  S3D & $704 \times 540$ 
  & $[422, 633] \times [270, 431]$ \\
  QMC & $69 \times 69 \times 115 $ 
  &  $[48, 68] \times [48, 68] \times [40, 74]$\\
  RTI & $144 \times 256 \times 256 $ 
  & $[74, 90] \times [66, 78] \times [88, 98]$\\
\midrule
 Dataset & Block B & Block C \\
 \midrule
CESM  & $[900, 1620] \times [450, 810]$
  & $[2159, 2879] \times [1079, 1439]$\\
  S3D
  & $[176, 387] \times [189, 350]$
  & $[70, 281] \times [28, 189]$\\
  QMC
  &  $[24, 44] \times [24, 44] \times [68, 102]$ 
  &  $[41, 61] \times [7, 27] \times [34, 68]$\\
  RTI
  & $ [57, 73] \times [23, 35] \times [156, 166]$ 
  & $ [44, 60] \times [51, 63] \times [204, 214]$ \\
  
  \bottomrule
  \end{tabu}
  }
  \vspace{-2mm}
  \caption{The sizes of four scientific datasets and the locations of their corresponding blocks.}
  \label{tab:blocks}
  \vspace{-4mm}
\end{table}
\section{Critical Point Extraction in the PL Setting}
\label{sec:PL-critical-points}

In the PL setting, Banchoff~\cite{banchoff1970critical} proposed a method to detect critical points, and Edelsbrunner and Harer~\cite{edelsbrunner2022computational} gave a detailed description as follows. 
Assuming the input data is modeled as a scalar field $f$ defined on a PL manifold $\Mspace$, the sublevel set of the data is $f^{-1}(-\infty, t]=\{x \in \Mspace \mid f(x) \leq t\}$ for some $t \in \Rspace$. 
A \emph{proper face} $\tau$ of a simplex $\sigma \in \Mspace$ is the convex hull of a non-empty, strict subset of vertices in $\sigma$. 
We call $\sigma$ a \emph{proper coface} of $\tau$ and denote it by \modify{$\tau<\sigma$}. 
The \emph{star} $St(v)$ of a vertex $v\in \Mspace$ is a set of its cofaces $St(v)=\{ \sigma\in\Mspace | v<\sigma \}$. 
The smallest complex that contains the star is the \emph{closed star} $\overline{St}(v)$. 
The \emph{link} $Lk(v)$ of $v$ is the set of simplices in the closed star but not in the star. 
The link can be thought as the boundary of a small neighborhood around $v$. 
The \emph{lower link} $Lk^-(v)$ is a set of simplices in the link whose vertices all have values smaller than $v$:
\begin{equation}
   Lk^-(v)=\{ \sigma\in Lk(v) | \forall u\in \sigma, f(u)<f(v) \}.
 \end{equation}
 Correspondingly, the \emph{upper link} $Lk^+(v)$ is a set of simplices in the link whose vertices have values larger than $v$. 
 Vertex $v$ is a \emph{PL regular point} if both $Lk^-(v)$ and $Lk^+(v)$ are simply connected. Otherwise, it is a \emph{PL critical point}. 

\begin{figure}[h]
\vspace{-2mm}
   \centering
   \includegraphics[width=0.7\linewidth]{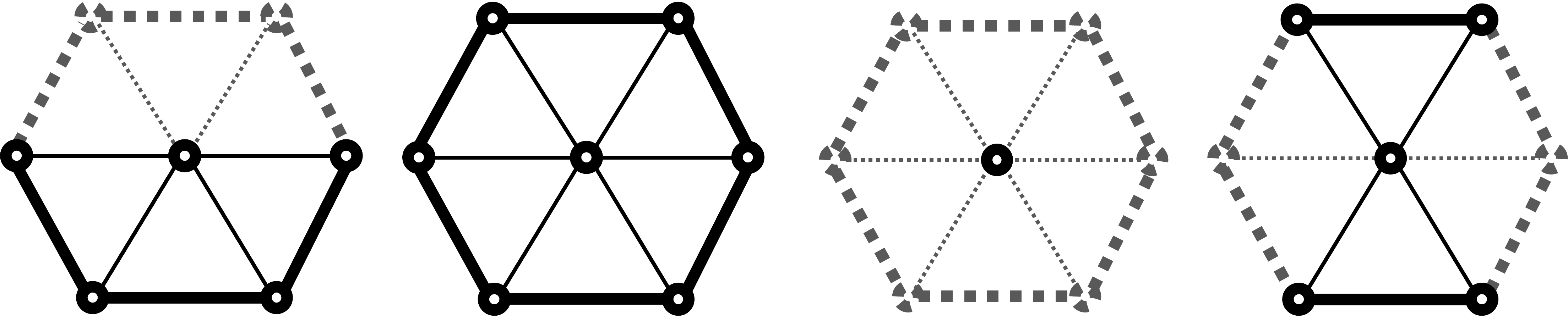}
   \vspace{-2mm}
   \caption{From left to right: the lower link of a regular point, a
   local maximum, a local minimum, and a saddle.}
 \label{fig:critical-point}
 \vspace{-2mm}
\end{figure}

 In a PL 2-manifold, there are three types of critical points: local minimum ($Lk^-(v)=\emptyset$), local maximum ($Lk^+(v)=\emptyset$), and saddles (when $Lk^-(v)$ or $Lk^+(v)$ is not simply connected). 
 We show examples of lower links in~\cref{fig:critical-point}. 
 For a PL 3-manifold, there are four types of critical points: local minimum, local maximum, and two different types of saddles.

\section{Pseudocode for Spatial Hashing}
\label{sec:spatial-hashing}

We provide the pseudocode for spatial hashing used to remove duplicated critical points in~\cref{sec:duplication-removal}.

\begin{algorithm}[H]
\caption{Removing Duplicated Critical Points}
\label{algorithm:spatial-hashing}
\begin{algorithmic}[1]
\REQUIRE 
A set of critical points $X$ and a threshold $\tau$.
\ENSURE
A set of critical points $Q$ after duplication removal.
\STATE $Y = \emptyset$, a duplicated set of critical points to be removed from $X$.  
\FORALL{$x \in X$}
\STATE $k=\frac{x}{20\tau}$
\STATE duplicate\_label = false
\FOR{$i \in [1,d]$}
\STATE ${\Ind}_{i,1} = \floor{k_i}$, ${\Ind}_{i,2}=\ceil{k_i}$
\ENDFOR
\STATE We obtain all hash indices in a set $\mathcal{I}=\{\mathrm{I} =({\Ind}_{1},\cdots,{\Ind}_{d})\}$ where each dimension $\Ind_i$ is chosen from ${\Ind}_{i,1}$ or ${\Ind}_{i,2}$. 
\FORALL{hash index $\mathrm{I} \in \mathcal{I}$} 
\STATE $V$ = boost::hash\_combine($\mathrm{I}$) \\
\FORALL{point $y$ in the hash buckets corresponding to  $V$}
\IF{$||x-y||<\tau$}
\STATE Identify $x$ as a duplicated critical point.
\STATE duplicate\_label = true
\STATE $Y = Y \bigcup x$.
\STATE Goto line 16, exit the two innermost for loops.
\ENDIF
\ENDFOR
\ENDFOR
\IF{duplicate\_label is false}
\STATE Register $x$ to all the hash buckets corresponding to $\mathcal{I}$.
\ENDIF
\ENDFOR
\STATE return $Q = X \setminus Y $
\end{algorithmic}
\end{algorithm}

\section{Detailed Complexity Analysis}
\label{sec:compplexity-details}

In span filtration, for every dimension's first derivative, $p(p+1)^{d-1}$ control points are utilized in every span. Let $n$ denote the number of spans. In the experiments, we consistently maintain $p\geq 2$. Within each span, it is necessary to evaluate all the utilized control points. The cost of span filtration is $\mathcal{O}(dp^dn)$. In Newton's method, the computation cost for every element in the Hessian matrix is $\mathcal{O}(p^d+dp^3)$. Each iteration of Newton's method involves computing the Hessian and gradient at $\mathcal{O}(d^2(p^d+dp^3))$ and solving linear system at $\mathcal{O}(d^3)$. Since $(p+1)^d$ initial points are used in each span, the cost of finding critical points in all spans is $\mathcal{O}(i_{max}d^2(p^{2d}+dp^{3+d})n)$. During spatial hashing, registration and comparison of at most $(p+1)^d$ points occur $2^d$ times per span. The spatial hashing takes $\mathcal{O}((2p)^ddn)$. In practical applications, $p$ and $d$ are always 2 or 3 which are much smaller than $n$ (and can be treated as constants). The complexity of span filtration and spatial hashing are smaller than finding all critical points via Newton's method.
The overall time complexity of CPE-MFA is $\mathcal{O}(i_{max}d^2(p^{2d}+dp^{3+d})n)$, which matches the time complexity of finding critical points in all spans. 
In all the experiments conducted, more than 95\% of the time was spent finding critical points using Newton's method, confirming the analysis that this step dominates the overall time complexity.


\end{document}